# Physics-Based Modeling of TID Induced Global Static Leakage in Different CMOS Circuits

Gennady I. Zebrev, Vasily V. Orlov, Maxim S. Gorbunov, Maxim G. Drosdetsky

*Abstract* — **Compact modeling of inter-device radiation-induced leakage underneath the gateless thick STI oxide is presented and validated taking into account CMOS technology and hardness parameters, dose-rate and annealing effects, and dependence on electric modes under irradiation. It was shown that proposed approach can be applied for description of dose dependent static leakage currents in complex FPGA circuits.**

## I. Introduction

The problem of radiation-induced leakage in CMOS circuits is a challenge which questions the main merit of the CMOS technology – its low consumption in the off-state regime [1]. It is well-known that the total ionizing dose (TID) induced supply current of the CMOS circuits is added from the two components. First, it is the intra-device edge drain-to-source leakage currents through the narrow conductive paths near the transistor sidewall isolation which is proportional to the number of fingers [2]. Second, it is the inter-device leakage through parasitic conductive paths under the thick Shallow Trench Isolation (STI) oxides between, for instance (see Fig. 1), the n+ source/drain region of an n-channel device and the n-well region of an adjacent p-channel device [3, 4].

There is a large body of experimental evidence that radiation-induced supply current is not proportional to the finger number, which implies the importance of inter-device leakage, especially for sub-100 nm technologies [5, 6, 7,]. This means that the IC supply leakage is not an additive sum of leakages in the separate local transistors, and it is more a global response of the whole circuit. One of the distinctive features of the effective parasitic transistor structure of inter-device leakage is a lack or remoteness of the conductive gate above thick STI oxide.

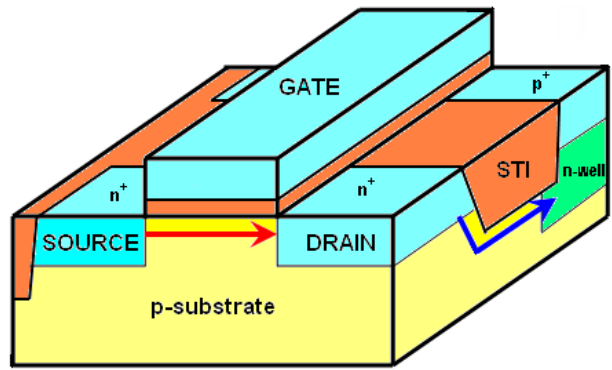

Fig. 1. Cross-sectional diagram indicating: (1) drain-to-source leakage and (2) leakage between the n+ source/drain region of an n-channel device and the n-well region of an adjacent p-channel device.

Despite the low values of electric fields $E_{ox}$ in such oxides (typically $\leq 10^5$ V/cm), they are capable to accumulate a significant number of positively charged defects near the interface between the isolation oxide and the p-Si substrate, causing occurrence of the parasitic electron channels [8, 9].

The objective of this study is to develop a physics-based analytical model for compact simulation of the dose dependencies of total IC supply current as functions of the total dose at different dose rates, assuming the dominance of inter-device component.

The rest of this paper is organized as follows. The Sec. II focuses on a concise description of the physics-based model and its features. The model validation and radiation-oriented applications are presented in Sec. III. Sec. IV is devoted to simulation of the TID induced circuit leakage in FPGAs.

## II. Parasitic leakage current modeling

### A. Electrostatics of the gateless transistor structure

The positive charge accumulated in the oxide leads to the surface potential changing underneath the thick oxide. The surface potential shift depends on the thickness of the oxide, concentration of charged traps and doping of the p-epi substrate. The electric neutrality condition without the gate is described by the equation $N_{ox} = N_S$. Here, $N_S$ is the total (negative) charge density in the silicon substrate, which can be represented using the charge-sheet approximation as a sum of the channel $n_S$ and the depletion layer densities [10]

$$N_S = N_A x_d(\varphi_S) + n_S, \qquad (1)$$

Manuscript received September 29, 2017.
G. I. Zebrev, V. V. Orlov, M. G. Drosdetsky are with the Department of Micro- and Nanoelectronics of National Research Nuclear University MEPHI, Moscow, Russia, e-mail: gizebrev@mephi.ru.
M. S. Gorbunov is also with Scientific Research Institute of System Analysis, Russian Academy of Sciences, Moscow, Russia.
This work was supported by the Competitiveness Program of the NRNU MEPHI.

where $x_d(\varphi_S)$ is the depletion layer width, $N_A$ is the Si substrate acceptor concentration. On other hand, we have from the Poisson equation solution [11]

$$N_S = N_A L_D \left[ \exp\left(\frac{\varphi_S - 2\varphi_F}{\varphi_T}\right) + \frac{\varphi_S}{\varphi_T} \right]^{1/2}, \quad (2)$$

where $\varphi_F = \varphi_T \ln N_A / n_i$ is the Fermi level position marker, where $\varphi_T = k_B T / q$ is the thermal potential, $n_i \cong (N_C N_V)^{1/2} e^{-E_G/2k_B T}$ is the silicon temperature-dependent intrinsic concentration ($\sim 10^{10}$ cm$^{-3}$ at 300K), $N_C$ ($N_V$) is the effective conduction (valence) band density, $E_G$ is the Si bandgap.

The Debye length is defined here as follows

$$L_D = \left(\frac{2\varepsilon_S \varepsilon_0 \varphi_T}{q N_A}\right)^{1/2}, \quad (3)$$

where $\varepsilon_S \varepsilon_0$ is the Si permittivity. Eq. 2 is valid on a condition $\varphi_S > \varphi_F$ when the electron inversion layer underneath the thick STI has already formed.

### B. Surface potential as functions of oxide trapped charge

Then the electric neutrality condition can be written a follows

$$\exp\left(\frac{\varphi_S - 2\varphi_F}{\varphi_T}\right) + \frac{\varphi_S}{\varphi_T} \cong a^2, \quad (4)$$

where $a \equiv N_{ox}/N_A L_D$. When one can neglect the exponential term in (4), i.e., at $2\varphi_F > \varphi_S > \varphi_F$ (depletion mode), we have

$$\varphi_S \cong \frac{q N_{ox}^2}{2\varepsilon_S \varepsilon_0 N_A}. \quad (5)$$

This as a usual form of the expression for surface potential has a quadratic dependence on concentration of the charge in the oxide [12]. The exact solution of this equation can be written generally in a following form

$$\varphi_S = \varphi_T a^2 - \varphi_T W\left[\exp\left(a^2 - \frac{2\varphi_F}{\varphi_T}\right)\right] = $$
$$= \frac{q N_{ox}^2}{2\varepsilon_S \varepsilon_0 N_A} - \varphi_T W\left[\frac{n_i^2}{N_A^2}\exp\left(\frac{q N_{ox}^2}{2\varepsilon_S \varepsilon_0 N_A \varphi_T}\right)\right], \quad (6)$$

where $W(s)$ is the Lambert function, defined as a solution of the equation $W(se^s) = s$, and also known as the ProductLog function in *Mathematica* [13].

Figure 2 shows the surface potential underneath the thick STI oxide as functions of external oxide charge, calculated with (6) at different doping levels of the p-type Si substrate.

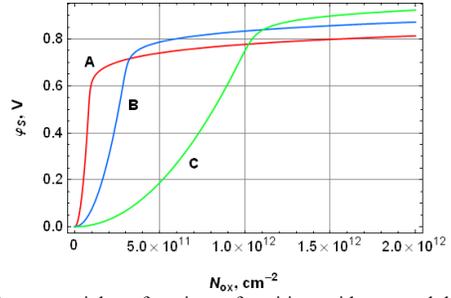

Fig. 2. Surface potentials as functions of positive oxide-trapped defects densities at different doping levels of the p-type substrate. (A) $N_A = 10^{15}$ cm$^{-3}$, (B) $N_A = 10^{16}$ cm$^{-3}$, (C) $N_A = 10^{17}$ cm$^{-3}$, the oxide thickness $t_{ox} = 500$ nm.

Note that additional substrate doping hinders the band bending at moderate $N_{ox}$ (i.e., at $N_{ox} < N_A x_D(2\varphi_F)$), since we have $\varphi_S \propto 1/N_A$ in (5), and, at the same time, it provides the larger maximum values of $\varphi_S$ because of an increase in $\varphi_F$.

### C. Electron density as a function of the oxide trapped charge

Then, the parasitic electron density as function of $N_{ox}$, $N_A$, temperature can be calculated using the charge-sheet approximation

$$n_S = N_S - N_A x_d(\varphi_S) = N_{ox} - N_A L_D (\varphi_S/\varphi_T)^{1/2} = $$
$$= N_{ox}\left\{1 - \left[1 - 2\frac{\varepsilon_S \varepsilon_0 N_A \varphi_T}{q N_{ox}^2} W\left(\frac{n_i^2}{N_A^2}\exp\left[\frac{q N_{ox}^2}{2\varepsilon_S \varepsilon_0 N_A \varphi_T}\right]\right)\right]^{1/2}\right\}. \quad (7)$$

This compact analytical formula with the clear physical parameters is the main result of this work.

General relations (6) and (7) have quite hermetic forms. To make them more physically transparent, it is instructive to consider some special cases. For a case $N_{ox} < N_A x_D(2\varphi_F)$, which equivalent to the depletion mode condition $\varphi_S < 2\varphi_F$, we have (5) and

$$n_S \cong \frac{n_i^2}{N_A}\frac{\varepsilon_S \varepsilon_0 \varphi_T}{q N_{ox}}\exp\left(\frac{q N_{ox}^2}{2\varepsilon_S \varepsilon_0 N_A \varphi_T}\right). \quad (8)$$

For the inversion case $\varphi_S \geq 2\varphi_F$ ($N_{ox} > N_A x_D(2\varphi_F)$) we have an asymptotic relation

$$n_S \cong N_{ox} - N_A x_d(2\varphi_F) \quad (9)$$

Figures 3 show the dependencies of the parasitic channel electron densities as functions of the oxide-trapped charge calculated at different substrate doping levels.

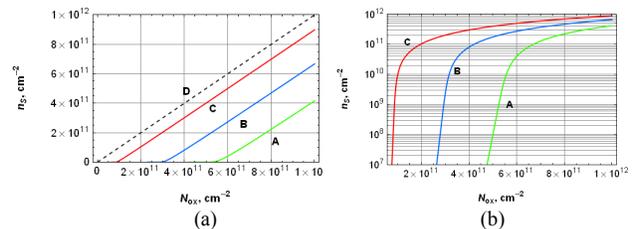

Fig. 3. Electron surface densities beneath the thick STI oxide in a linear (a) and logarithmic (b) scales as functions of $N_{ox}$ densities at the different doping levels of the p-type substrate: (A) $N_A = 10^{15}$ cm$^{-3}$, (B) $N_A = 10^{16}$ cm$^{-3}$, (C) $N_A = 3\times 10^{16}$ cm$^{-3}$, (D) ultimate curve $n_S = N_{ox}$, $t_{ox} = 500$ nm.

Notice that the calculated dependencies of $n_S$ on $N_{ox}$ have a threshold form which is caused by the deep physical reasons. Until the surface potential $\varphi_S$ reaches the value $\varphi_F$, the electron conductive channel beneath the oxide is absent in principle. So, the substrate doping effectively suppresses the formation of parasitic electron channels under the oxides. A formal inclusion of the interface traps in this model does not significantly change the qualitative and even quantitative results but introduces additional uncertainties associated with the inability to determine their parameters.

### D. Current off-state leakage as a function of the oxide trapped charge

In fact, we are going to model the leakage current of the parasitic field oxide field effect transistors (FOXFET) with the thick STI oxide as a gate dielectric but without the gate itself. In contrast to usual I-V characteristics, we are interested in simulation of the leakage currents as functions of the oxide-trapped charge and, implicitly, as functions of the ionizing dose. The simulated parasitic FOXFET is assumed to be in a saturation mode. Thus, the leakage saturated current $I_L \cong I_{DSAT}$ can be written as a sum of the diffusion and the drift components [14, 15]

$$I_L \cong \frac{W}{L} q \mu_0 \varphi_T n_S + \frac{W}{2L} \frac{\mu_0 q^2 n_S^2}{C_D(\varphi_S)}, \quad (10)$$

where $W/L$ is the width to length ratio of the parasitic transistor structure, $\mu_0$ is electron mobility, the depletion layer capacitance is calculated in a standard way

$$C_D(\varphi_S) = \frac{\varepsilon_S \varepsilon_0}{x_d(\varphi_S)} = \frac{\varepsilon_S \varepsilon_0}{L_D} \left( \frac{\varphi_T}{\varphi_S} \right)^{1/2}. \quad (11)$$

The first term in (10), corresponding to a linear dependence of the drain current on $n_S$, is the diffusion current, dominating at low $n_S$. The second term, dominating at high $n_S$, corresponds to the saturated drift current in the square-law approximation. Leakage current is expressed in (10) as an explicit function of the surface potential which in turn calculated as a function of the accumulated oxide charge. Notice that the threshold voltage $V_T$ and the oxide capacitance $C_{ox}$ are both missing in this model, which describes the case of the remote gate, i.e., when $C_{ox} \ll C_D(\varphi_S)$.

## III. I-V MODEL VALIDATION

### A. Dose effect modeling

Taking into account the tunnel annealing, the buildup of the oxide-trapped on the Si-SiO$_2$ interface can be estimated as follows [16, 17]

$$N_{ot} = \eta_{eff} F_{ot} K_g t_{ox} D \left( 1 - \frac{\lambda}{\ell} \ln \left( \frac{D}{P t_1} \right) \right), \quad (12)$$

where $P$ is a dose rate, $t_{ox}$ is the STI oxide thickness, $\eta_{eff}$ is the effective charge yield, $F_{ot}$ is the dimensionless hole trapping efficiency, $K_g \cong 8 \times 10^{12}$ cm$^{-3}$rad (SiO$_2$)$^{-1}$ is the electron-hole pair generation rate constant in SiO$_2$, $\ell$ is the effective width of the oxygen vacancy precursors for the oxide hole traps, $\lambda$ is the minimum tunnel length ($\leq 0.1$ nm), $t_1$ is a reference time. For simplicity, we will keep the same value $\lambda/\ell = 0.05$ and $t_1 = 0.1$ s in all simulation in this paper.

### B. Simulation at different interface characteristics

Dose-dependent leakage underneath the thick STI is quite sensitive to the hole trapping efficiency and the electric-field-dependent charge yield in the oxides. For instance, Fig. 4 shows the dose dependencies simulated with the equations (7-12) at different $F_{ot}$.

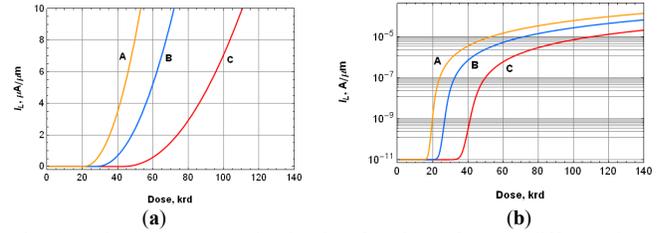

Fig. 4. Leakage currents $I_L$ simulated as functions of TID at different charge trap efficiencies in a linear (a) and logarithm (b) scales. (A) $F_{ot} = 0.04$, (B) $F_{ot} = 0.03$, (C) $F_{ot} = 0.02$: $W/L = 1$, $t_{ox} = 500$ nm, $\mu_0 = 300$ cm$^2$/(V×s), $\eta_{eff} = 0.5$, $N_A = 3 \times 10^{15}$ cm$^{-3}$, $P = 100$ rd(Si)/s.

The existence of "the dose threshold" for the leakage current is typical for most of the experimental data [18, 19]. In practice, the dose threshold can be caused by a high level of the dark (non-radiation-induced) supply current of the circuit, which generally cannot be modeled with the proposed approach.

### C. Simulation at different operation temperatures

An increase in the supply current at elevated temperatures is an important problem, especially for the modern highly scaled circuits. Figure 5 shows the total-dose dependencies of circuit leakage current simulated at different operation temperature. It was assumed that the temperature dependence of the mobility in the usual manner $\mu_0(T) = \mu_0 (T_0/T)^{3/2}$ [11].

Like the subthreshold region of usual MOSFET's I-V characteristics [20] the leakage current increases with temperature at low doses due to the Boltzmann statistics of non-degenerate electrons in parasitic channels.

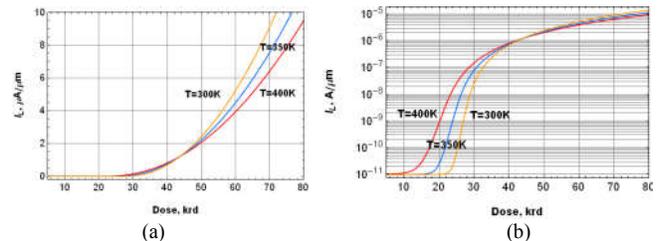

Fig. 5. Supply currents as functions of TID at different operation temperatures in a linear (a) and logarithm (b) scales. $F_{ot} = 0.03$, $W/L = 1$, $t_{ox} = 500$ nm, $\mu_0 = 300$ cm$^2$/(V s), $\eta_{eff} = 0.5$, $N_A = 3 \times 10^{15}$ cm$^{-3}$, $P = 100$ rd(Si)/s.

Meanwhile, the supply leakage in the inverted at relatively high doses channels slightly decreases at elevated operation temperatures due to the temperature degradation of electron's mobility.

## D. Validation at different dose rates

Figure 5 shows typical dependence of supply current as a function of dose in comparison with experimental data [21] at different dose rates.

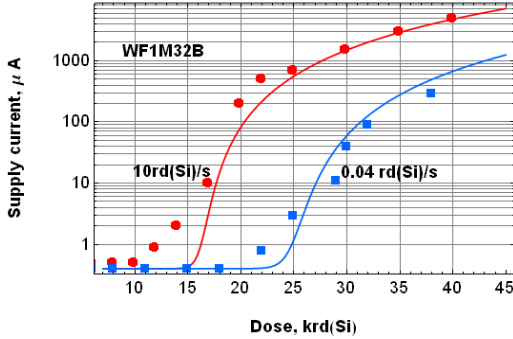

Fig. 5. Comparison of simulation (solid lines) experimental data for irradiated WF1M32B CMOS circuits [21] at dose rates 10 rd(Si)/s (circles) and 0.04 rd(Si)/s (squares). Fitting parameters: $N_A = 3 \times 10^{15}$ cm$^{-3}$, $W/L = 1000$, $t_{ox} = 500$ nm, $\mu_0 = 400$ cm$^2$/(V×s), $F_{ot} = 0.09$, $\eta_{eff} = 0.3$.

The dose curves in Fig. 5 exhibit reducing degradation at lower dose rates, typical for all CMOS devices. This is caused by the time-dependent form of the tunnel relaxation in (12). The true dose-rate effects (i.e., ELDRS) can be incorporated into the computational scheme as an explicit decreasing dependence of the charge yield on an instant value of a dose rate, typical for the thick bipolar oxides [22].

## E. Validation at different electric modes

Figure 6 shows a comparison of simulated and experimental data at different electric biases taken from [23]. The charge yield $\eta(E_{ox})$ in (10) was modeled by a simplified empirical expression [3]

$$\eta_{eff}(E_{ox}) = \eta_0 + \frac{E_{ox}/E_0}{1 + E_{ox}/E_0}(1 - \eta_0), \quad (13)$$

where fitting parameters $\eta_0 = 0.05$ and $E_0 = 0.15$ MV/cm were used for simulation.

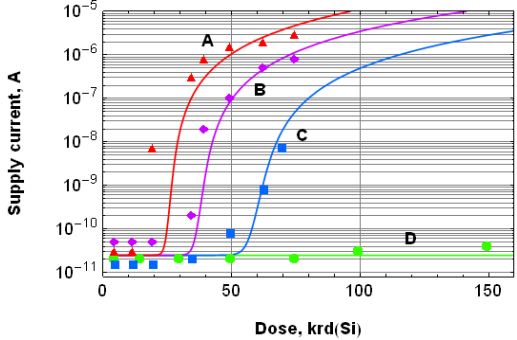

Fig. 6. Comparison of simulation and experiment for 180 nm test CMOS structures irradiated at a dose rate $P \sim 80$ rd(Si)/s at different bias conditions [23]:(A) $V_{GS} = 3.3$ V (triangles); (B) $V_{GS} = 1.8$ V (rhombuses); (C) $V_{GS} = 0.9$ V (squares); (D) $V_{GS} = 0$ V (circles); dose rate $P = 80$ rd(Si)/s. Fitting parameters: $t_{ox} = 400$ nm, $N_A = 2 \times 10^{15}$ cm$^{-3}$, $W/L = 1$, $\mu_0 = 300$ cm$^2$/(V×s), $F_{ot} = 0.04$.

As can be seen from this figure, the leakage currents in CMOS devices are very sensitive to the electric field in thick isolation during irradiation and such sensitivity can be well explained by a standard empirical approximation like (13). As usual, the worst-case irradiation bias is the ON state [24].

## IV. FPGA DOSE DEPENDENT STATIC LEAKAGE SIMULATION

Field-Programmable Gate Arrays (FPGAs) are the reconfigurable integrated circuits based on a high logic density regular structure, which can be customized by the end user via programmable switches of different types (SRAM, Antifuse, and FLASH) for the realization of different designs [25]. FPGAs are widely used for space applications due to their high flexibility and low-cost [26, 27]. However, the significant radiation-induced degradation of the static supply current is one of the major factors limiting a usage of FPGAs in radiation harsh environment [28]. We intend to show in this work that the proposed model of the chip-level leakage current allows us to describe the radiation-induced leakage degradation in the different types of FPGAs in a unified manner.

The proposed model has been validated with the literature experimental results for different types of FPGAs. In essence, we fitted here the parameters of a global parasitic transistor contributing to the total supply current change under ionizing irradiation.

For example, Fig. 7 shows a comparison between simulation and experiment for the Antifuse-based FPGA fabricated with the 0.25 μm technology process [18].

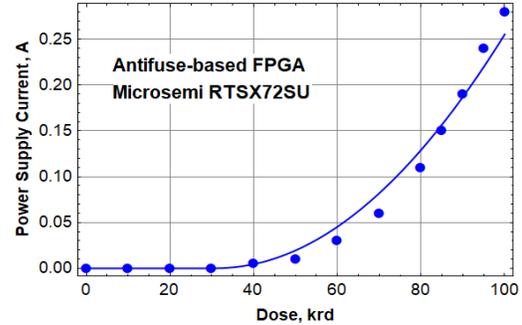

Fig. 7. The comparison between simulation results and experimental data for Antifuse-based FPGA RTSX72SU static supply current change during irradiation at $P = 16.7$ rd(Si)/s. Model parameters: $F_{ot} = 0.065$, $W/L = 1 \times 10^4$, $N_A = 5 \times 10^{15}$ cm$^{-3}$, $d_{ox} = 300$ nm, $I_{off0} = 0.023$ mA.

The value of the pre-rad leakage current $I_{off0}$ is determined mainly by the logic blocks in such FPGAs, and, thus, turns out to be relatively low [29, 30]. As can be seen in Fig. 7, the total supply current increases during irradiation by several orders of magnitude. Note that the fitting aspect ratio value of the effective parasitic transistor was rather high in this case. This is typical for all simulated complex circuits.

Figs. 8 and 9 show the supply current dose dependencies for the two SRAM-based FPGAs: 0.35 μm Xilinx XC4036XL [31] and 0.22 μm Xilinx XQVR300 [32].

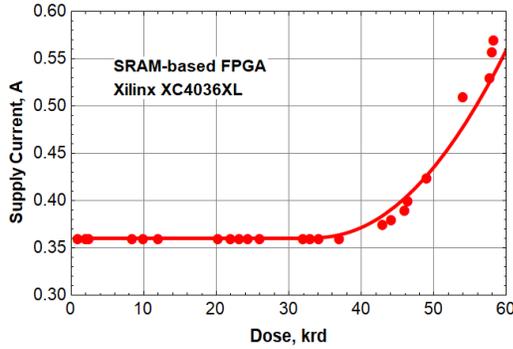

Fig. 8. The comparison between simulation results and experimental data for the SRAM-based FPGA XC4036XL static supply current change during irradiation at $P = 0.133$ rd(Si)/s. Model parameters: $F_{ot} = 0.039$, $W/L = 9 \times 10^4$, $N_A = 3 \times 10^{15}$ cm$^{-3}$, $d_{ox} = 500$ nm, $I_{off\,0} = 360$ mA.

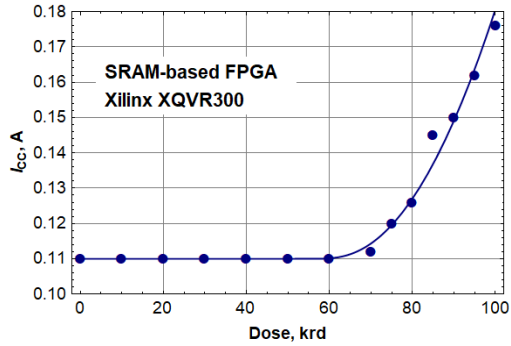

Fig. 9. The comparison between simulation results and experimental data for SRAM-based FPGA XQVR300 static supply current change during irradiation at $P = 50$ rd(Si)/s. Model parameters: $F_{ot} = 0.033$, $W/L = 3 \times 10^4$, $N_A = 5 \times 10^{15}$ cm$^{-3}$, $d_{ox} = 300$ nm, $I_{off\,0} = 110$ mA.

Note a similarity of the leakage current dose dependencies in both SRAM-based FPGAs. In contrast to the Antifuse-based FPGA, the supply current increases here only by several times. This is due to a high level of pre-irradiation supply current in these circuits.

Fig. 10 shows the comparison between the simulation results and experimental data for the highly scaled FLASH-based FPGA Microsemi RTG4 (65 nm technology node) [33]. Two important points can be noted in this case. First, this circuit has a higher threshold dose and TID hardness. Second, the fitting aspect ratio of the "global parasitic transistor" was relatively large. Simulation results and fitting suggest that the former point is likely due to the thinner isolation oxide used in a 65 nm process (see Eq.12). The latter point is caused in our opinion by a low technology node size. Actually, the effective width of the circuit parasitic transistor should of an order of the circuit linear size which is practically technology node independent. At the same time, the effective length of the parasitic transistor should be correlated with the technology node size.

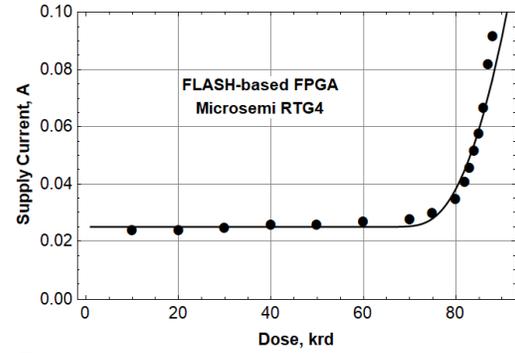

Fig. 10. The comparison between simulation results and experimental data for FLASH-based FPGA RTG4 supply current change during irradiation at $P = 167$ rd(Si)/s. Model parameters: $F_{ot} = 0.097$, $W/L = 1.8 \times 10^5$, $N_A = 7.5 \times 10^{15}$ cm$^{-3}$, $d_{ox} = 100$ nm, $I_{off\,0} = 25$ mA

Then the aspect ratio is $W/L \sim 1$ cm/50 nm $= 2 \times 10^5$. This numerical estimation well confirms the concept of the global parasitic transistor.

## V. Conclusion

Radiation-induced inter-device leakage is simulated using the physics-based analytical model. Comparison with the experimental results are presented to validate the model proposed in this paper. We have found also that such approach can be successfully used for the supply current radiation-induced degradation in FPGA circuits. Thus, despite an apparent particular form of the physical model, the validation results suggest that this generic approach can be used to describe the radiation-induced leakage currents in a wide range of CMOS devices, including complex circuits. Such unexpected efficiency of the particular model of the leakage currents underneath the thick oxides can be explained by a universal role of the thick isolation in advanced microelectronics. This paper is based on a report, presented at RADECS 2017 [34].

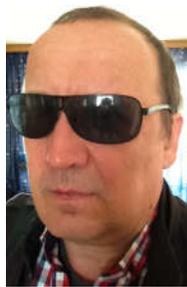

**Gennady I. Zebrev** was born in Tambov Province, Russia. He graduated from the department of Theoretical Physics of Moscow Engineering Physics Institute (MEPHI) in 1984. From 1984 to 1999, he was a senior researcher at the Research Institute of Science Instruments, Lytkarino, Moscow region. Since 1999 he has been an Associate Professor and hereafter a Full Professor at the Department of Micro- and Nanoelectronics of National Research Nuclear University MEPHI. He received the PhD in 2003 and the Doct. of Sci. degrees in technology in 2009. His research interests include device physics and modeling, general physics of nonequilibrium processes, physics and modeling of radiation effects in devices and nanoscale integrated circuits. He is the author of three books, several chapters in books and more than 90 articles. He was awarded in 2007 by the M. V. Keldysh medal of merit of Russian Space Federation and in 2016 by the Russian Government Prize in Education for the book "Physical Basics of Silicon Nanoelectronics" (2008, 2011).